\documentclass{article}
\usepackage{ijcai22}

\usepackage{times}

\usepackage{graphicx,subfigure}
\usepackage{subfigure}
\usepackage{soul}
\usepackage{url}
\usepackage[hidelinks]{hyperref}
\usepackage[utf8]{inputenc}
\usepackage[small]{caption}
\usepackage{graphicx}
\usepackage{amsmath}
\usepackage{amsthm}
\usepackage{booktabs}
\usepackage{algorithm}
\usepackage{amssymb}
\usepackage{algorithmic}
\usepackage{xcolor}
\def\draft{0}
\newcommand{\chaidi}[1]{{\if\draft1\color{blue}\fi#1}}

\title{Secure Forward Aggregation for Vertical Federated Neural Networks}

\author{
Shuowei Cai$^{1}$
\and
Di Chai$^{1,2}$\and
Liu Yang$^{1,2}$\and
Junxue Zhang$^{1,2}$\and
Yilun Jin$^{1}$\and
Leye Wang$^{3}$\and
Kun Guo$^{4}$\And
Kai Chen$^{1}$
\affiliations
$^1$
iSING Lab, The Hong Kong University of Science and Technology\\
$^2$Clustar
$^3$Peking University\\
$^4$Fuzhou University
\emails
scaiak@cse.ust.hk,
dchai@cse.ust.hk,
lyangau@cse.ust.hk,
jzhangcs@cse.ust.hk,
yilun.jin@connect.ust.hk,
leyewang@pku.edu.cn,
gukn@fzu.edu.cn,
kaichen@cse.ust.hk
}

\begin{document}

\maketitle

\begin{abstract}  % effectiveness
\chaidi{Vertical federated learning (VFL) is attracting much attention because it enables cross-silo data cooperation in a privacy-preserving manner. While most research works in VFL focus on linear and tree models, deep models (\textit{e.g.}, neural networks) are not well studied in VFL. In this paper, we focus on SplitNN, a well-known neural network framework in VFL, and identify a trade-off between data security and model performance in SplitNN. Briefly, SplitNN trains the model by exchanging gradients and transformed data. On the one hand, SplitNN suffers from the loss of model performance since multiply parties jointly train the model using transformed data instead of raw data, and a large amount of low-level feature information is discarded.  On the other hand, a naive solution of increasing the model performance through aggregating at lower layers in SplitNN (\textit{i.e.,} the data is less transformed and more low-level feature is preserved) makes raw data vulnerable to inference attacks. To mitigate the above trade-off, we propose a new neural network protocol in VFL called Security Forward Aggregation (SFA). It changes the way of aggregating the transformed data and adopts removable masks to protect the raw data. Experiment results show that networks with SFA achieve both data security and high model performance.
}

% On the one hand, SplitNN suffers from the loss of model effectiveness since multiply parties jointly train the model using transformed data instead of raw data and a large amount of low-level feature information is discarded. On the other hand, naively solution of increasing the model effectiveness through aggregating at lower layers in SplitNN (\textit{i.e., the data is less transformed and more low-level feature is preserved}) will leak more private data.

\end{abstract}

%The performance gap between the federated model and the centralized model is due to security protection. The trade-off between data security and model performance is complex and troublesome. In this paper, we focus on SplitNN, a neural network framework in vertical federated learning (VFL) that splits a complete network model for several participants. SplitNN reduces the interaction between models held by different participants and trains them by exchanging gradients and transformed data. However, this way of exchanging information makes it suffer from both performance loss and security risks. The transformed data of the passive party exposed contains abundant information about the raw data and is vulnerable to attacks. However, defending against attacks in SplitNN can easily lead to degradation of model performance. Based on these results, we design Security Forward Aggregation (SFA) protocol to mitigate this trade-off. It changes the way to aggregate the transformed data and generates removable masks for it to protect the raw data. Experiment results show that network with SFA achieve both raw data security and high model accuracy. 
\section{Introduction}

% Federated learning (FL) \cite{advanceFL} is \chaidi{a new paradigm of} collaborative machine learning with privacy preservation, and it \chaidi{could be} categorized into several categories according to different data partition \chaidi{scenarios}\cite{flbook}. Among them, vertical federated learning (VFL) \chaidi{is defined as} the scenario \chaidi{in which multiple} participants hold \chaidi{the same entities but } different features. There are a large number of mature and practical data models, such as logistic regression \cite{vlr,secure-bilevel} and decision trees\cite{sbt} in VFL. However, current works mainly focus on linear models and trees, while research on complex models like neural network is relatively insufficient. There is little analysis about the data security and model performance in complex models.
\begin{figure}[h]
	\centering
	\includegraphics[scale=0.36]{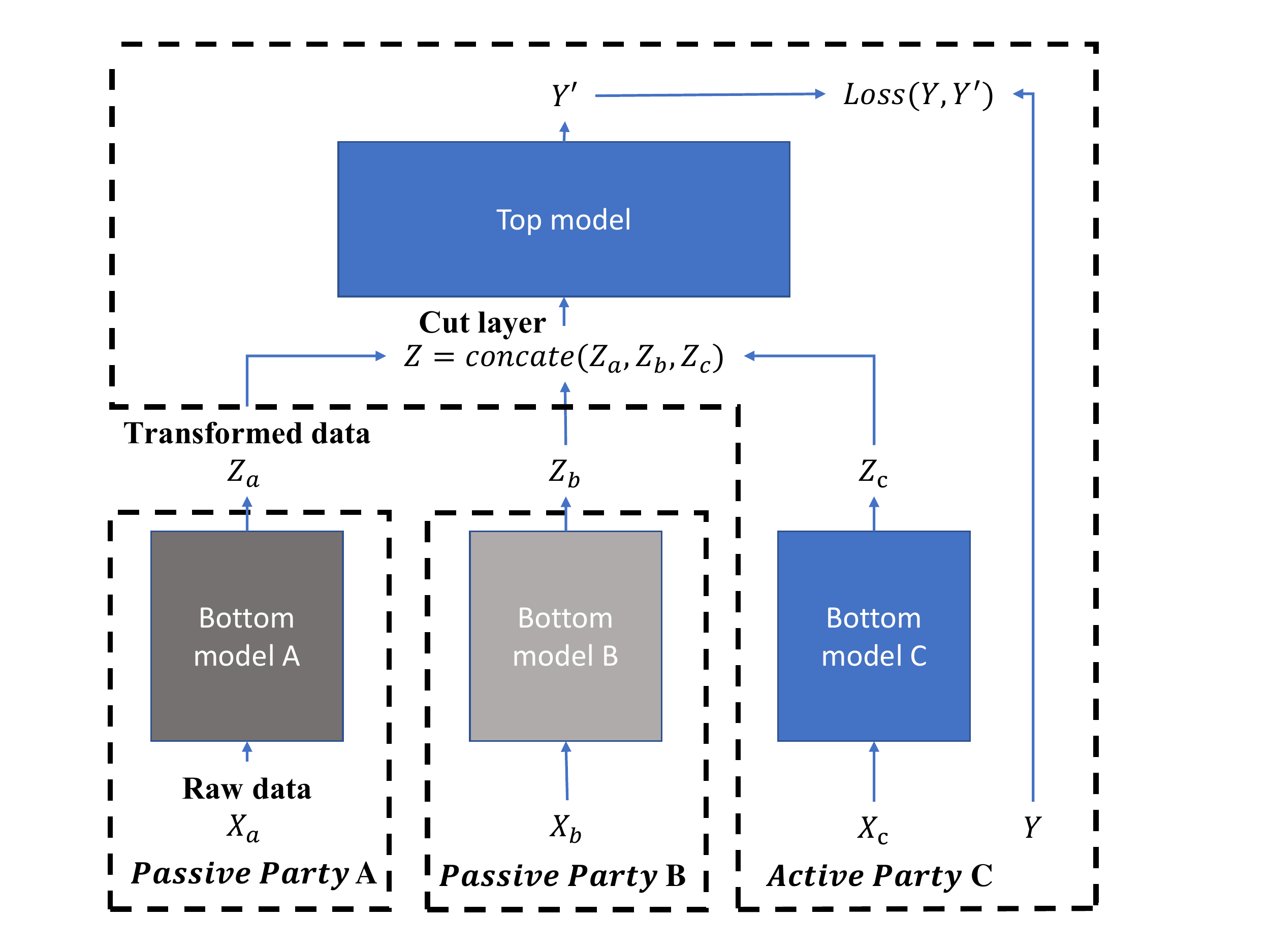}
	\caption{Split neural network in VFL}
	\label{fig:SplitNN}
\end{figure}
Federated learning (FL) \cite{advanceFL} is a new paradigm of collaborative machine learning with privacy preservation, and it could be categorized into several categories according to different data partition scenarios \cite{flbook}. Among them, vertical federated learning (VFL) is defined as the scenario in which multiple participants hold the same entities but different features. Existing work has explored various types of VFL models, such as logistic regression \cite{vlr,secure-bilevel} and decision trees \cite{sbt}. However, the neural network is not well studied in the VFL scenario. Specifically, there is little analysis of data security and model performance in complex models like neural networks.

% 1.FL作用 2. FL->VFL， 3. VFL简单模型 ->复杂模型 4.SplitNN 
%然而这些都是相对简单的线性模型，能力有限且不足以解决复杂任务。而在复杂任务中，人们需求强大的模型，如VFL中的神经网络，splitNN。
%

Split neural network (SplitNN) \cite{split_learning_for_health} is a framework able for neural networks in VFL. As shown in Fig.\ref{fig:SplitNN}, the whole network is partitioned into a top model and several bottom models. Each VFL participant keeps a bottom model for transforming its raw data, and the transformed data are passed to the top model at the cut layer to make the prediction in each forward process. However, exchanging the transformed data between participants result in a trade-off between data security and model performance in SplitNN.

%Different from the centralized neural network, the bottom models in splitNN are separated, and the hidden units inside a participant's bottom model can only receive information from the features own by itself. This special design is to reduce the information leakage of the raw data. However, it also imports a trade-off between data security and model performance due to the transformed data exchanged between participants.

% \chaidi{Existing work of adopting neural networks in VFL brings trade-off} between data security and model performance. In this paper, we mainly focus on one the most popular NN studies in FL, which is Split neural network (SplitNN) \cite{split_learning_for_health}. \chaidi{And VFL is one of the most important application scenarios of SplitNN.} \chaidi{Fig.\ref{fig:SplitNN} illustrate the framework of SplitNN}, the whole network is partitioned into a top model and several bottom models. Each VFL participant keeps a bottom model for transforming its raw data, and the transformed results are passed to the top model at the cut layer for further analysis in each forward process. Different from the centralized neural network, the bottom models in \chaidi{S}plitNN are separated, and the hidden units inside a participant's bottom model can only receive information from the features own by itself. This special design \chaidi{reduces} the information leakage of the raw data. However, it also imports a trade-off between data security and model performance due to the transformed data exchanged between participants.

On the one hand, using the transformed data instead of the raw data for prediction will result in a loss of model performance because the transformed data contains only a part of the information in the raw data. According to the analysis in \cite{understanding_image}, a significant amount of low-level feature information is discarded while the raw data passes through the layers. This discarding information happens while the raw data passes through the bottom models of SplitNN, and it also transforms data from fine grain to coarse grain. For example, using the term "cheap baby products" to represent a classic cotton diaper from a famous brand. However, this transformation increases the difficulty of the model's prediction because the model cannot capture the interaction between the discarded information,  just like the classic "beer and diaper" relationship in data mining. A person who buys beer may be interested in diapers, but it's hard to say that a person who buys a drink will be interested in baby products. When two participants in SplitNN possess feature information like beer and diapers, low-level feature interaction loss happens, and it will decrease the model's performance.

%It is because the hidden units in a bottom model can not receive any feature information from other parties, and the hidden units in the top model know nothing about the discarded low-level feature information. Therefore, SplitNN is unable to capture the low-level feature interactions between participants. In fact, the missing connections between the bottom models are supposed to fulfill this purpose. Thus, the discarded information and the lack of low-level feature interaction result in a performance loss in SplitNN.

%However, the discarded information also protects the raw data, which makes the exposure of the bottom model's output secure. But no previous work has evaluated the information leakage from the transformed data and its performance against attacks. Also, the degree of discarded low-order information may affect both raw data security and model accuracy. In our experiments, we find a trade-off between security and model performance of SplitNN: the more profound the data is transformed, the more secure the raw data is, but the worse the model will perform, and vice versa.
On the other hand, the transformed data which are sent directly to the active party leaks information about the raw data. The passive party's raw data will be vulnerable to inference attacks if the transformed data contains too much information about it. As a result, it is inappropriate to increase the amount of information in the transformed data to improve the model's performance. Thus, we identify this trade-off, as the transformed data influences both the model performance and data security. It is impossible to achieve high model performance and high data security simultaneously in SplitNN.

Motivated by the above problem, we propose a method to mitigate this trade-off between data security and model performance. We consider that the direct exposure of the transformed data is improper, and the data protection by controlling the amount of information in the transformed data is hazardous. To this end, we proposed a Secure forward aggregation (SFA) protocol that can securely aggregate the bottom models' output without exposing the individual output from the bottom model. We modify the aggregation method at the cut layer and provide a removable mask to protect the passive party's transformed data and ensure raw data security. With SFA, we mitigate the trade-off and can achieve lossless performance compared to the centralized model with high security.
%Inspired by \cite{CEASAR}.

The main contributions of this paper are summarized as follows:
\begin{itemize}
\item We evaluate the trade-off between the model performance and the security of raw data in SplitNN in vertical federated learning.
\item We present a Secure forward aggregation protocol to protect the participant's transformed data while being lossless. With SFA, we can mitigate this trade-off and achieve both good model performance and high data security in neural networks in VFL.
\end{itemize}

\section{Motivation}
In this section, we first introduce splitNN, one of the most popular frameworks of neural networks in VFL. While noting the special designs for VFL in this architecture, we also analyze the trade-off between data security and model performance.

\subsection{Background: SplitNN in VFL}
As shown in Fig.\ref{fig:SplitNN}, splitNN is a distributed network structure in VFL that support multiparty settings. The participants of SplitNN are categorized into the active party (participant with labels) and the passive party (participant without labels). Each passive party holds one bottom model for local data transformation. The active party holds both a bottom model and a top model and uses the top model to make predictions with the transformed data from all participants. 

\begin{figure}[h]
	\centering
	\includegraphics[scale=0.42]{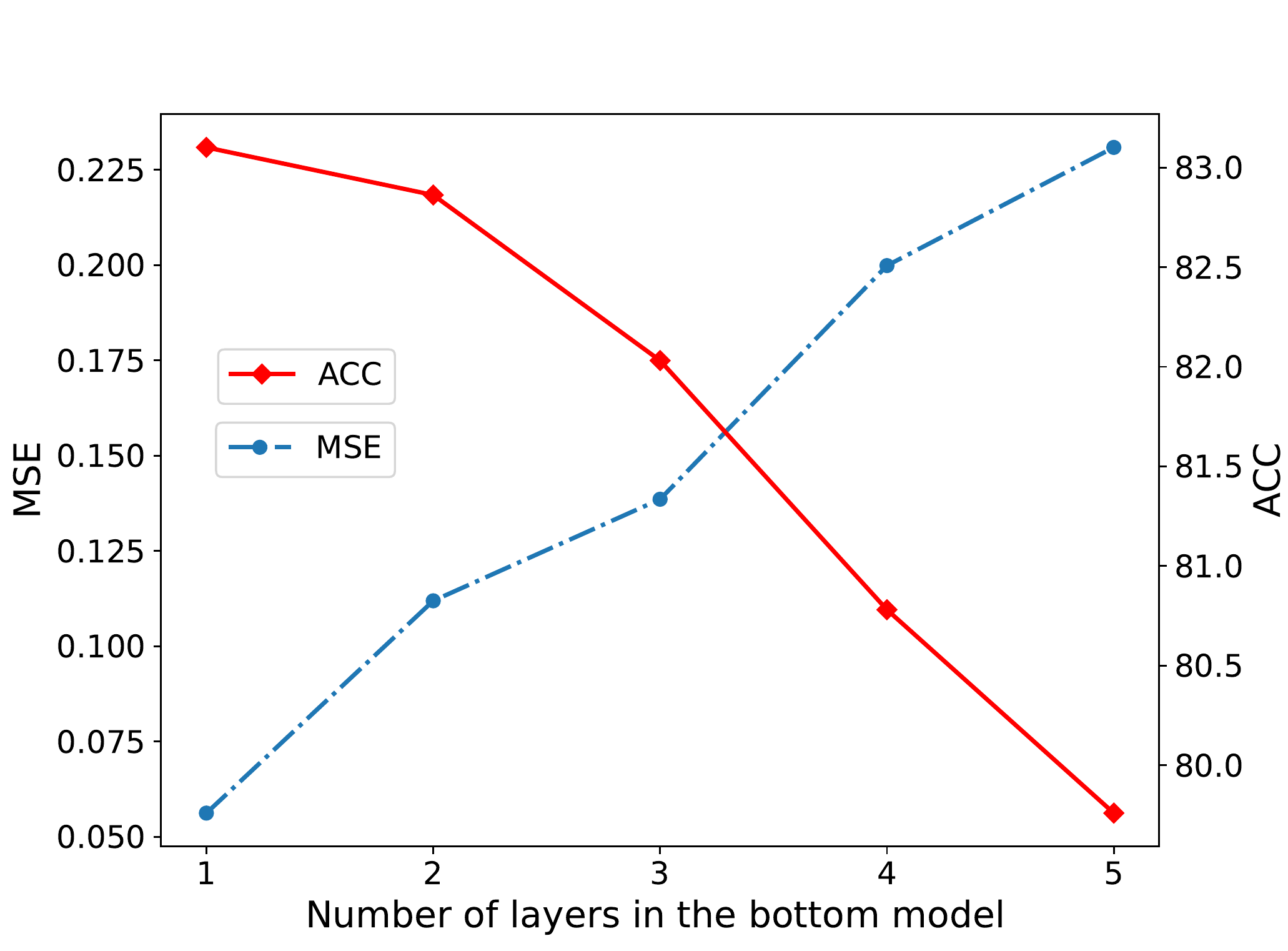}
	\caption{Performance-security trade-off in splitNN}
	\label{fig:trade-off}
\end{figure}

The forward process of SplitNN consists of three steps: 1) participants use their bottom models to transform the data, 2) passive parties send the transformed data to the active party, and 3) the active party will concatenate these transformed data, applies activation functions and feed it into the top model to get the prediction results. The layer which concatenates these transformed data is also called the cut layer. In the backward process, the active party will first update the top model normally and calculate the gradients of the embeddings. Then it will send the gradients of these embeddings to their owners to update their bottom models.

%着重讲一下top model的concat & 训练时的梯度下降和上传。

\subsection{Trade-Off of SplitNN}
The participant's raw data security is a primary concern in vertical federated learning. In SplitNN, one substantial information leakage is the transformed data directly sent to the active party. Those transformed data do not contain all raw data information because some low-level information is lost during the forward propagation. However, there is still a correlation between the transformed and original data. When the active party of SplitNN applies inference attacks like \cite{feature_inference_attack} to dig out the correlation, the raw data is able to be approximated.

Indeed, discarding more information from the transformed data is a way to improve data security in SplitNN. Increasing the number of layers in the bottom model is a feasible way to fulfill this. A higher bottom model will increase the complexity of the transformation and discard more low-level feature information from the transformed data, making raw data hard to reconstruct. However, the discarded low-level feature information in the transformed data is crucial to the model's performance. It contributes to the model performance in low-order feature interactions between participants. The connections between the bottom models are used to capture these low-level feature interactions, but they are missing in SplitNN.

Therefore, to further measure the relationship between data security and model performance in SplitNN, we use the generative regression network (GRN) in \cite{feature_inference_attack} to attack the SplitNN and generate approximation data to get close to the raw data. As shown in Fig.\ref{fig:trade-off}, the approximation data of GRN is far from the raw data when the number of layers in the bottom model is high, but the model performance drop significantly(\textit{i.e.,} 3\%). The model's performance is the best when there is only one layer in the bottom model. However, the restored data is closest to the raw data in this case, and the MSE between the approximation data and the raw data is only 0.055. Model performance and raw data security become the two ends of the scale in SplitNN. As a result, we urgently need a method to mitigate this trade-off.

%我们在使用ICDE的这篇文章的GRN来对embedding进行还原，并且不断调整bottom model的高度来观察其还原的接近成果以及此高度下模型的准确率。从图二中可以看出，当我们降低bottom model的高度时，模型的准确度逐步提升，但对应的还原数据更接近真实，这就导致了一个安全性问题。而当模型的高度能够保证还原数据的结果接近于随机猜测时，某些数据集上的模型甚至会有10%的损失，这就导致了一个模型performance的问题。因此，我们迫切的需要一个能够缓和或解决这个trade-off的办法。
%

%额外问题：1. 这么定义这个损失是否存在问题?
%2. 协议的安全性分析应该写在哪里?
%3. 如何提及FDML的实验结果？

\section{Secure Forward Aggregation}
\subsection{Overview}
In this section, we propose a novel protocol called Secure Forward aggregation (SFA) to protect the transformed data of the passive parties. It provides removable masks to passive parties, and the masks do not introduce noise into the computation of the model. SFA protocol is used at the topmost layer for the bottom models of all participants. It securely aggregates the bottom models' output values without exposing their true values using a summation operation with masks. In the mask generation of SFA, we securely share a part of the transformed data from the active party using homomorphic encryption and send it to the passive party. The shared result will be the mask that protects the passive party's output, and homomorphic encryption ensures that the active party knows nothing about the value of the mask. Therefore, the mask effectively protects the passive party's raw data without introducing noise into the training. 

Different from methods like secure aggregation \cite{secure_agg}, secure forward aggregation can protect the passive party's input in the aggregate output even in the two-party scene. In SFA, we use a weight mask generated by the passive party and sent to the active party under homomorphic encryption to produce masks for the transformed data. The weight mask is seen as a part of the weight of the topmost layer of the active party's bottom model and is also used to prohibit the active party from knowing the actual output of its bottom model. Therefore, the active party cannot recover the passive party's input from the aggregated result. Moreover, SFA could be applied to multi-participant scenarios, allowing all passive parties to keep masks to protect their transformed data. With SFA, we can aggregate the transformed data securely regardless of the information it contains. Therefore, the trade-off between model and security is moderate, and we can train a model that both performs well and protects raw data perfectly.

\begin{figure}[h]
	\centering
	\includegraphics[scale=0.35]{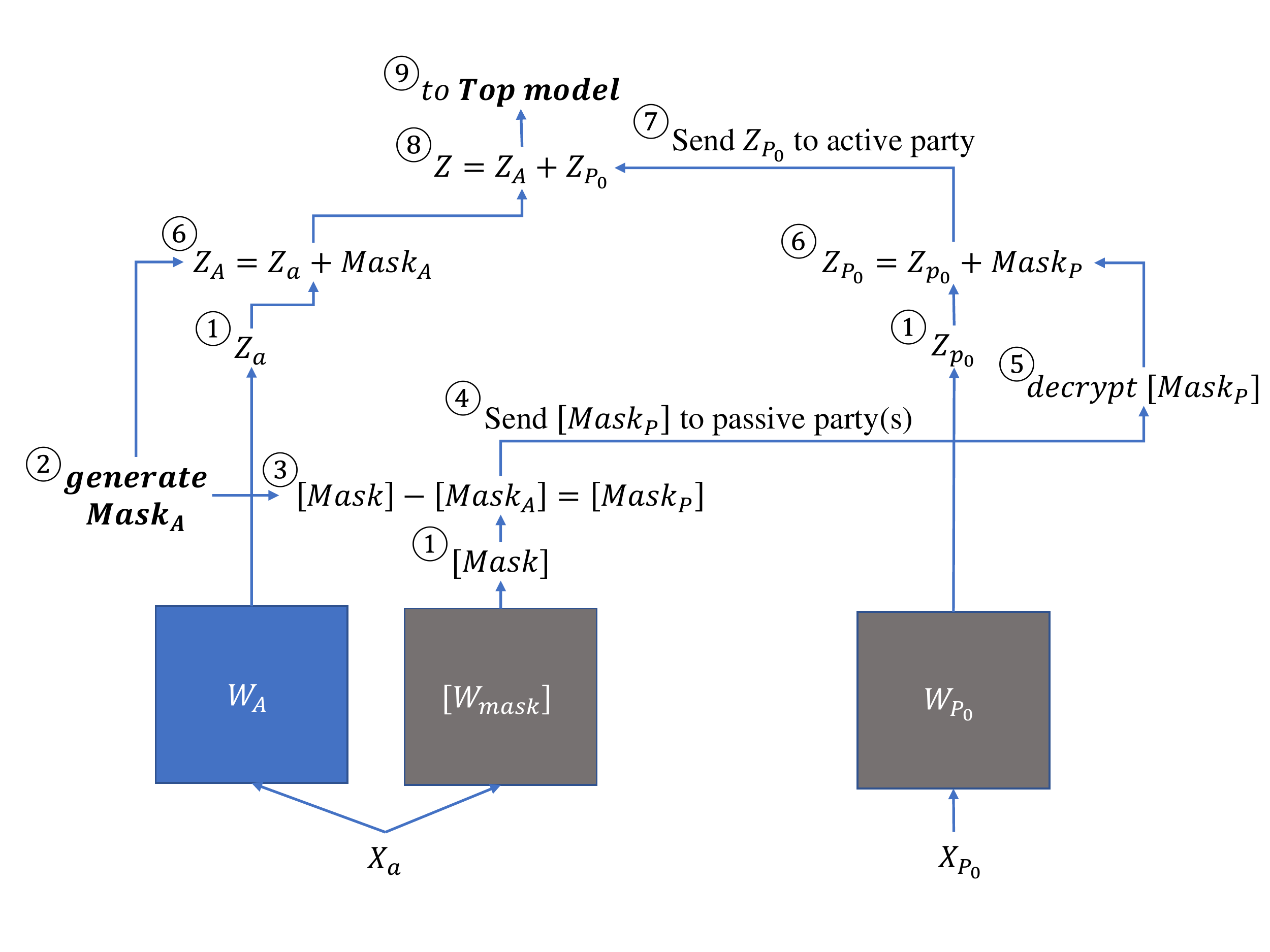}
	\caption{Secure Forward Aggregation in two party scene}
	\label{fig:SFA}
\end{figure}

\subsection{Aggregation Method}
We sum the transformed data from the bottom models in SFA instead of concatenating them in SplitNN. The first reason is that the concatenate operation exposes the transformed data directly, increasing the difficulty of data protection. Moreover, the concatenate operation treats each neuron independently, but sum will not. So, the feature interaction will not be captured at the cut layer. Therefore, concatenation operation at the cut layer will indirectly increase the loss of data information by passing the transformed data through another layer for capturing feature interaction. As a result, we change the aggregation method at the cut layer from concatenating operation to a summation operation.

\subsection{Training with Weight Mask}
The algorithm of the Secure Forward Aggregation protocol is shown in algorithm\ref{alg:algorithm}. During the initialization stage, one passive party will generate a weight mask ($W_{mask}$ in algorithm\ref{alg:algorithm}) and send it to the active party as a part of the weights in the topmost layer of the bottom model. This weight mask will be encrypted by Paillier homomorphic encryption \cite{paillier} and sent to the active party. This weight mask will never be decrypted, and it will be used to generate masks for passive parties without letting the active party know.

Fig.\ref{fig:SFA} shows the forward process and steps of secure forward aggregation in a two-party setting. $A$ and $P_0$ are the active and passive parties; $X_a$ and $X_{p_0}$ here are the raw data or the values output from hidden units; $W_A$ and $W_B$ here are (part of) the weight of the topmost layer of the bottom model, $[W_{mask}]$ is the weight mask, and $[\cdot]$ represents homomorphic encryption. In the forward process of SFA, both parties will calculate $Z_a$ and $Z_{p_0}$ normally using $W_A$,$X_a$ and $W_{P_0}$,$X_{P_0}$. Then, the active party will calculate $[Mask]$ using the encrypted value $[W_{mask}]$. Then, it will generate a random matrix $mask_A$ and subtract it from the mask and send the remaining part $[Mask_{P}]$ back to the passive party $P_0$. Then, $P_0$ will decrypt the result and obtain its mask. After that, the two parties add the mask onto their transformed data, and the passive party $P_0$ sends its masked transformed data to the active party for aggregation. Finally, the active party will sum these transformed data to obtain the final result $Z$. 

Secure Forward Aggregation protocol can also protect the transformed data in a multiparty setting. If there is extra passive party other than party $P_0$, party $P_0$ will continue share the masks to other passive parties, calculate $Mask_{P_0}= Mask_{P}-\sum_{i=1}^n Mask_{P_i}$ and send the new mask $Mask_{P_i}$ to passive party $P_i$. In this way, the active party still knows nothing about the masks.

\subsection{Removable Mask on Transformed Data}
Unlike methods that follow differential privacy to generate noise to protect their intermediate result, the mask generated in the SFA protocol is a part of the original output. Therefore, the mask in SFA will not introduce noise to the aggregated value. It is because we regard the encrypted $W_{mask}$ as a part of the weight for the active party, and the actual weight for the last layer in the bottom model of the active party should be $W_{A_{true}}=W_{mask}+W_{A}$. It is clear to see that the final output $Z$ is the sum of the two party's transformed data:

\begin{equation}
\begin{aligned}
    Z & =Z_A+\sum_{i=0}^N Z_{P_i}=Z_a+Mask_A+\sum_{i=0}^N (Z_{P_i}+Mask_{P_i})\\
    & =(W_{A}+W_{mask})X_a+\sum_{i=0}^N W_{P_i}X_{P_i}\\
    & =W_{A_{true}}X_a+\sum_{i=0}^N W_{P_i}X_{P_i}
\end{aligned}
\end{equation}

We can see that $W_{A_{true}}X_a+\sum_{i=0}^N W_{P_i}X_{P_i}$ are the aggregated result of all transformed data. The masks for passive parties are a part of $W_{A_{true}}X_a$. They are generated by the encrypted weight mask $[W_{mask}]$ and shared using a random matrix, and added back to the final aggregated result. In the backward process, the participants calculate the gradient normally, and all parties will add the gradients to the plaintext weight. During the whole training process, $W_{mask}$ is kept unchanged. Therefore, we can consider it a noise initially added to the model weights. As the model is updated continuously, the impact of this noise will gradually fade away when doing gradient descends and updating the weights of the plaintext part.

%多参与方场景也是能做的。当partyB接收到noise矩阵后，只需要将mask再进行分发，将结果发给别的参与方做mask即可。这样新生成的mask也不会对结果造成任何影响，而且actvie party对其一无所知。

\begin{algorithm}[tb]
\caption{Secure forward aggregation}
\label{alg:algorithm}
\textbf{Participants Settings}: Active party $A$, Passive party $P_0$, ($\{P_i|i=1\dots n\}$ for other passive party in multiparty scene)\\
\textbf{Input}: Batch of raw data or embedding from hidden units hold by participants of VFL: $X_a$,$X_{p_0}$ ($X_{p_i}$ for other passive parties),\\
%\textbf{Initialization}: weight matrix $W_A$\& $W_B$ for party A \& B\\
\textbf{Output}: Aggregated result $Z$\\
\textbf{Initialization}
\begin{algorithmic}[1] %[1] enables line numbers
\STATE Active party $A$ generates weight matrix $W_{A}$, Passive party $P_0$ generates weight matrix $W_{P_0}$. (If there are other passive parties, they generate their own weight matrix $W_{p_i}$)\\
\STATE Party $P_0$ generates HE key pair $\{sk_b, pk_b\}$. Generate weight matrix $W_{mask}$ , encrypt it using $sk_b$ and send the encrypted result $[W_{mask}]$ to party $A$.
\end{algorithmic}
\textbf{Forward process} 
\begin{algorithmic}[1]
\STATE All parties obtain the next batch of data (or hidden units value from the layer below)
\STATE Party $A$ calculates $Z_a=W_{A}X_a$ and $[Mask]=[W_{mask}]X_a$, Party $P_0$ calculates $Z_{p_0}=W_{P_0}X_{P_0}$. (Other passive parties calculate $Z_{p_i}=W_{P_i}X_{P_i}$)
\STATE Party $A$ generates random matrix $Mask_A$, calculate $[Mask_{P}]=[Mask]-[Mask_A]$ and send it to party $P_0$.
\STATE Party $P_0$ decrypt $[Mask_P]$. (If there are other passive parties, $P_0$ generates random matrix $Mask_{P_i}$ and send it to party $P_i$ and calculate: $Mask_{P_0}=Mask_P-\sum_i^n Mask_{P_i}$)
\STATE Party $A$ calculates $Z_A=Z_a+Mask_A$, Party $P_0$ calculates $Z_{P_0}=Z_{p_0}+Mask_{P_0}$ and send it to party $A$ (other passive parties calculate $Z_{P_i}=Z_{p_i}+Mask_{P_i}$ and send it to party $A$)

\end{algorithmic}
\textbf{Backward process} 
\begin{algorithmic}[1]
\STATE Active party send the upper gradient to each participants
\STATE All participants use this gradient to update their bottom model. $[W_{mask}]$ in party $A$ is kept unchanged.
\end{algorithmic}
\end{algorithm}

\subsection{Security Analysis}
This subsection discusses the security of Secure Forward Aggregation in a semi-honest setting, which is the standard security assumption in federated learning. We show that the transformed data are well protected in the Secure Forward Layer, and the passive parties cannot infer the data from the active party.

\subsubsection{Passive Party's Data Security}
In the SFA protocol, the transformed data $Z_{p_i}$ for passive party B is protected by a mask $Mask_{P_i}$. The active party will only obtain the masked result, and it cannot distinguish the mask and the transformed data from the masked result. Even though it knows that $Mask$ is a transformation of $X_b$, there are infinite eligible values of $Mask$. Therefore, it is insufficient to infer the exact value of mask $Mask_{P}$ or $Mask_{P_i}$ and to further infer the passive party's raw data. 

\subsubsection{Active Party's Data Security}
The active party's transformed data is secure because they are not sent outside. Though the passive party knows that $Mask$ is a transformation of $X_b$, it knows nothing about the random generated $Mask_A$. Therefore, the passive party cannot infer $Mask$ to perform further inference attacks, and the active party's data security is ensured.

\subsection{Mitigate Trade-off using SFA}
We have already shown that SFA ensures the security of the transformed data. Therefore, we can keep a shallow bottom model for better performance. When there is only one fully-connected layer in the bottom model, and the aggregate method is changed from concatenation to summation, the structure of the model is the same as the centralized neural network. Therefore, the performance degradation caused by the model architecture no longer exists.

Though the weight mask also impacts training, with reasonable settings, the initialized weights of the weight mask will not significantly impact the final results of the model. We initialize the weight mask using a uniform distribution bounded by $ 2/\sqrt{in\_features}$ and encrypt it, then send it to the active party. The weight generation of this weight mask follows \cite{linear_init}, and it reduces the impact of the weight mask of model training and the final performance.

%需要注意的是，maskB在
%当我们把SFA放在第一层时，我们发现，由于底层输入不会随着模型的更新而变动，模型的mask便可以复用以避免重复计算。当使用这个技巧时，我们可以在训练开始前预先生成mask，然后

\section{Experiment}
In the experiment sections, extensive experiments are done to show how SFA can mitigate the trade-off between data security and model performance. 

%讲我们用了什么数据集，他们的来源？我们怎么处理这些数据集的
\begin{table}
\centering
\begin{tabular}{lllll}
\toprule
Datasets  &  Sector  &News20  & Amazon  & FMNIST \\
\midrule
Datasize   & $9619$ &  $18,846$  & $100,000$   & $60,000$  \\
Features     & $55,197$ & $173,762$  & $257$ & $784$       \\
Labels       & $105$ & $20$   & $2$  & $10$  \\
\bottomrule
\end{tabular}
\caption{Datasets descriptions}
\label{tab:datasets}
\end{table}

\begin{table}
\centering
\begin{tabular}{lrrr}
\toprule
Datasets  & Sector  &News20  &  Amazon  \\
\midrule
Centralized  &  $91.28_{\pm 0.39}$  & $83.63_{\pm 0.27}$ & $77.46_{\pm 0.10}$     \\
SplitNN     & $86.43_{\pm 0.51}$  &$79.76_{\pm 0.68}$ &  $72.86_{\pm 0.07}$       \\
SFA-NN(ours)       & $90.86_{\pm 0.30}$   &$83.86_{\pm 0.32}$ &  $77.44_{\pm 0.10}$    \\
\bottomrule
\end{tabular}
\caption{model performance on different datasets (ACC) }
\label{tab:performance}
\end{table}

%讲我们做了什么实验，要从什么角度来分析这个问题。
\subsection{Experiment Setting}
We use a neural network structure with six fully-connected layers to illustrate the performance of SFA on neural networks in VFL. We select SplitNN and a centralized model (all data are integrated for modeling) to compare a neural network with SFA (SFA-NN). We also fixed the number of hidden units of each layer for fair comparison and ran the experiment of model performance for ten trials to reduce randomness in training.

We fix the dropout to 0.3, batch size to 256, and apply batch normalization to train the model for 50 epochs in default. Then, we select the best learning rate from \{$1e-1,1e-2,1e-3,1e-4,\dots$\} with zero regularization coefficient for all experiments. The default participant number of VFL is set to two, and the features are partitioned equally for each participant. The bottom model height is set to 5 for SplitNN and 1 for SFA-NN in default for a fair comparison with the same level of security.

\subsection{Dataset}
We use four classification datasets to demonstrate the performance problem and the trade-off in SplitNN: Sector \cite{libsvm}, news20 \cite{news20}, Amazon electronic \footnote{\url{http://jmcauley.ucsd.edu/data/amazon/}} and Fashion MNIST (FMNIST) \cite{fmnist} dataset. 

We preprocessed these data to meet the requirement of the experiments of SplitNN. We used the TF-IDF algorithm to transform the news20 data into a sparse matrix for training. We use a trained Deep Interest network \cite{din} to preprocess and transform 100,000 items in the Amazon electronic data into embeddings of 257 and treat them as data in model training. FMNIST is the dataset we demonstrate the security concerns in SplitNN, so we normalize the ranges of all feature values in it into (0, 1) as \cite{feature_inference_attack} for better demonstration.

The detailed descriptions of the data set after preprocessing are shown in table \ref{tab:datasets}.

\begin{figure}[h]
	\centering
	\includegraphics[scale=0.5]{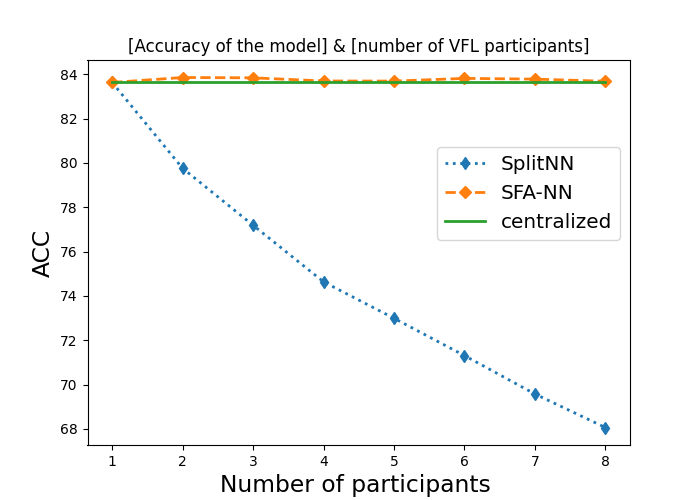}
	\caption{Model performance in multiparty settings }
	\label{fig:multiparty}
\end{figure}

\subsection{Performance of SFA}

We experiment and compare the performance of SplitNN and SFA-NN to show that our proposed method achieves good performance in high security. We also use the Centralized model in this experiment, which refers to a model with a standard neural network structure trained by pooling all data together in a non-federal learning setting. Because the performance of federated learning models should be as close as possible to the model performance in the non-federal settings \cite{flbook}, we use it as the target of SFA-NN to evaluate its performance. This baseline can precisely reflect the ability of SFA-NN to reduce the performance gap between neural networks in VFL and centralized neural networks.

\subsubsection{Model Performance under Two-Party Setting}
Table.\ref{tab:performance} shows the experiment result of SFA-NN in the two-party setting. The performance of SFA-NN is close to the centralized model and is significantly better than SplitNN. Although the weight mask of SFA impacts the model's training, it will not have a significant impact on the final performance of the model. The performance gap between SFA-NN and the centralized model is small, and for comparison, SplitNN's performance is low on these tasks, and SFA-NN's performance is significantly better than SplitNN.

\subsubsection{Model Performance under Multiparty Setting}
One of the reasons that SplitNN has gained popularity is that it supports multiparty training conveniently. Pessimistically, the more participants there are, the more feature partitions between participants will result in more low-level feature interaction loss. Thus, a severe model performance decrease will happen in the multiparty scenario of SplitNN. As shown in Fig.4, the model performance drops dramatically on the News20 dataset when the number of participants increases. But there is no such performance loss in SFA-NN, which shows that our method is effective in multiparty settings.

\subsection{Trade-off between Security and Model Performance}
In this experiment, we fix the total number of network layers and the number of neurons in each hidden layer. We then adjust the height of the cut layer and the height of the bottom model to observe the performance of the model and the security of the raw data. (the height of the top model decreases with the increase of the height of the bottom model and vice versa)

\begin{figure}
\centering     %%% not \center
\subfigure[model performance]{\label{fig:a}\includegraphics[width=41mm]{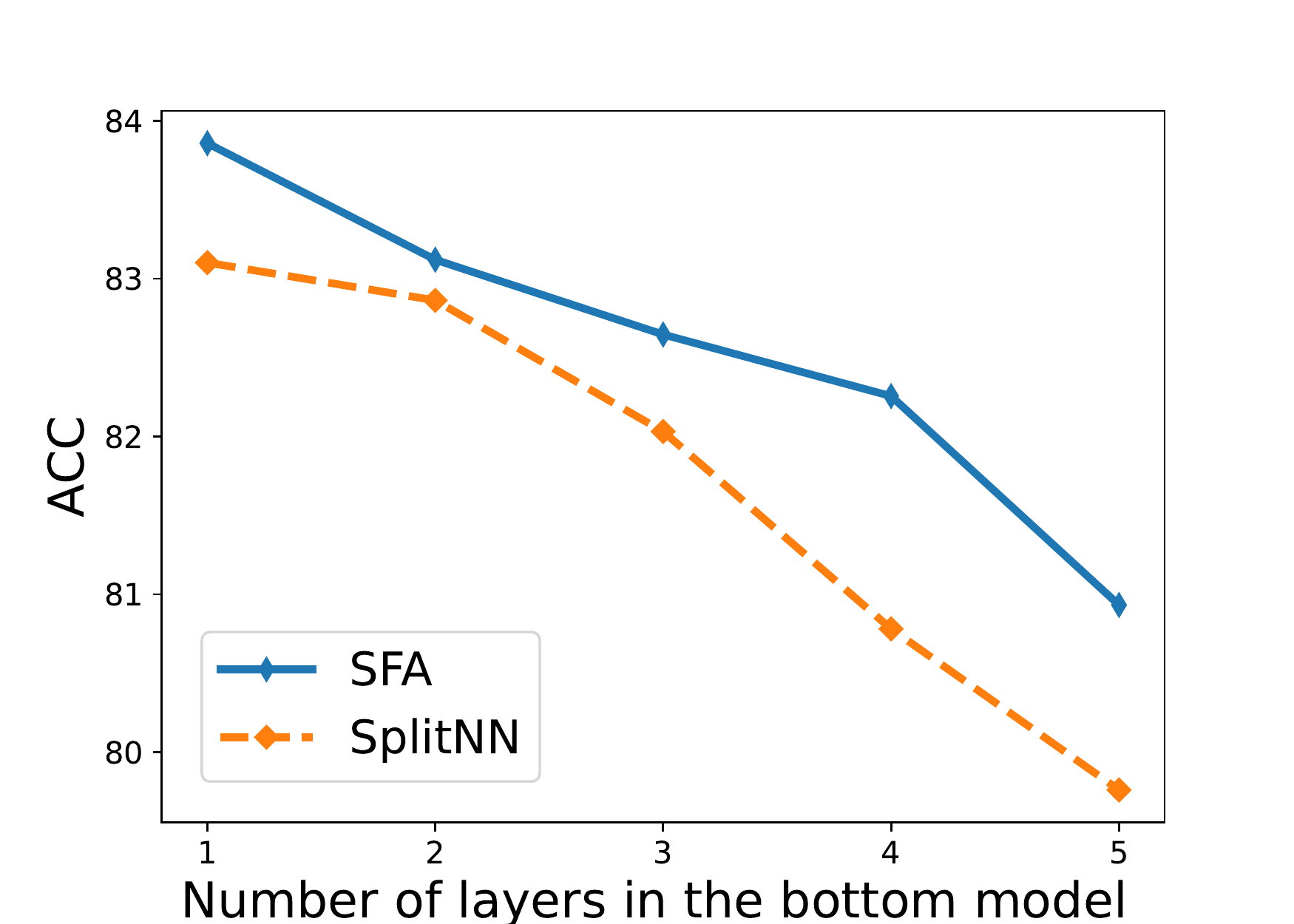}}
\subfigure[data security]{\label{fig:b}\includegraphics[width=41mm]{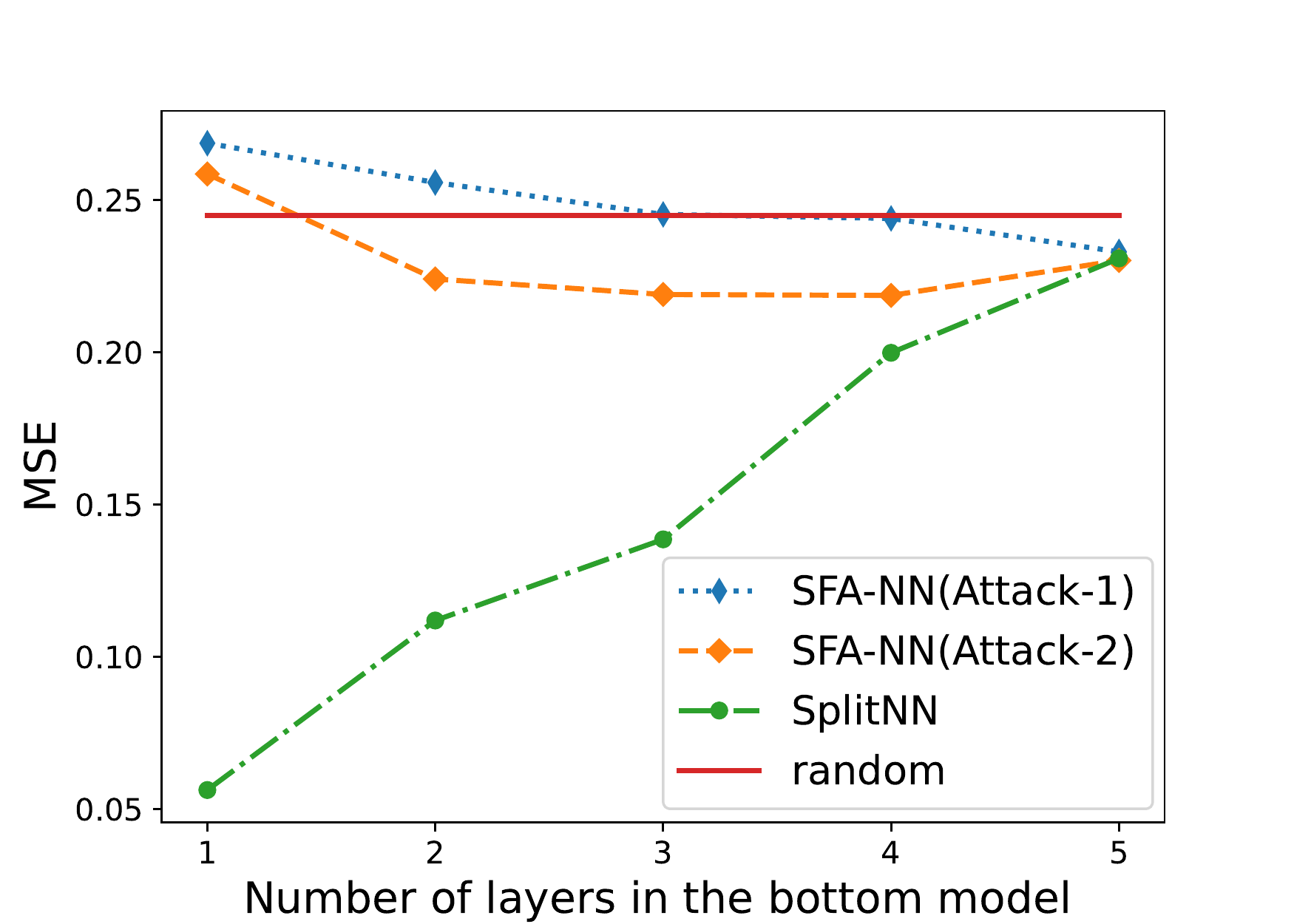}}
\caption{Trade-off between data security and model performance}
\label{fig:exp_trade_off}
\end{figure}

Fig.\ref{fig:a} shows the model performance on the News20 dataset with the bottom model of different heights. When the height of the bottom model increase, the model performance of SplitNN drops gradually. SFA-NN also suffers from this performance loss. Though the summation operation for aggregation at the cut layer improves the model performance, the model performance of SFA-NN is only similar to SplitNN with one less layer in the bottom model. The performance problem due to the discarded information has not been fundamentally solved. This experiment shows the damage that excessive discarding of low-level information brings to the model performance. Reducing the number of layers will be an intuitive solution for those seeking higher model performance, but this brings threats to the raw data.

To evaluate the information leakage of the transformed data, we train models using FMNIST datasets to 88\% accuracy with the length of the transformed data set to 256. Then, we use the generative regression network(GRN) \cite{feature_inference_attack} to attack the bottom model and reconstruct the raw data using the test dataset. GRN is the network to generate approximation data to approximate the passive party's raw data. We take the active party's data features and the passive party's transform data as the input to train the model. We train the GRN by minimizing the Mean Square Error(MSE) between the real transformed data and the transformed result of the approximation data. Because the transformed data $z_{p_0}$ are not known by the active party in SFA-NN, we use two masked outputs, $Z_P$ (attack-1) and $Z_P+mask_A$ (attack-2), to substitute the transformed data, and the protections of these two attacks are $Mask_{P_0}$ and $Mask$ respectively. We also use random values between 0-1 as a baseline of the attack to evaluate the attack method's performance and show the effectiveness of SFA's protection.

Fig.\ref{fig:b} shows the attack result on the transformed data, and the MSE metric indicates the distance of the attack results from the raw data. We can see that the attack is effective on the transformed data when the number of layers in the bottom model is low. When the layer number increase, the effect of the attack decrease, but the model performance gets lower. However, the attack is ineffective when SFA is used. GRN cannot achieve a good approximation of raw data even if the bottom model has only one layer. In fact, the MSE distance of the approximation data always gets larger as the training proceeds when the two attack methods act on the SFA, suggesting that the attack on transformed data with SFA is infeasible. In conclusion, SFA can protect the raw data with low bottom model layers. We can use SFA to gather the information from multiple participants at a low layer of the neural network and improve the model's performance.

% \begin{figure}[htbp]
% 	\centering
% 	\begin{subfigure}{0.49\linewidth}
% 		\centering
% 		\includegraphics[width=1\linewidth]{figure/exp3.png}
% 		\caption{model performance}
% 		\label{tradeoff:performance}
% 	\end{subfigure}
% 	\begin{subfigure}{0.49\linewidth}
% 		\centering
% 		\includegraphics[width=1\linewidth]{figure/exp4.png}
% 		\caption{data security}
% 		\label{tradeoff:security}
% 	\end{subfigure}
% 	\caption{Trade-off of security and model performance}
% 	\label{fig:trade_off}
% \end{figure}

% \begin{figure}[h]
% 	\centering
% 	\includegraphics[scale=0.45]{figure/exp4.png}
% 	\caption{Approximation error of the reconstruction data}
% 	\label{fig:security_height}
% \end{figure}

%我们知道我们在时间性能上表现不好，但我们会优化的

\section{Related Work}
FDML \cite{fdml} is another framework that supports neural networks in feature-partition settings with privacy-preservation. In FDML, each participant has an independent local model, and the final predictions of the model are obtained by summing the outputs of all local models. However, there is no direct connection between the local models of FDML, so it also suffers from a similar performance loss in SplitNN. This performance loss is also reflected in their experiments on neural networks. 

Moreover, due to the design assumption that labels are shared among participants in FDML, researchers seldom focus on label security in this framework. In fact, the gradient at the top layer of FDML exposes the labels directly \cite{label_attack}. Leakage of data labels is unacceptable for vertical federal learning. Therefore, we do not include it as a baseline in VFL.

%They also pointed out that the missing connections between the participants led to the model performance degradation, which is the same as our argument.
\section{Conclusion}
This paper proposes a Secure Forward Aggregation protocol to mitigate the trade-off between model performance and data security in SplitNN in VFL. This protocol provides removable masks to protect the transformed data in SplitNN and aggregates the information from different parties better. Experimental results show that we achieve almost the same performance as the centralized model, and we can keep the raw data safe and resistant to attacks using SFA. We effectively mitigate the trade-off between model performance and data security in neural networks in VFL.

This work still has some limitations. On the one hand, SFA introduces partial homomorphic encryption to perform secure computations, increasing the computational effort. Nevertheless, there are ways to reduce time consumption. For example, we can reduce the multiplication calculation of the same ciphertext weight masks and plaintext data and accelerate computation using parallelism and hardware\cite{haflo}. On the other, there is a lack of hyperparameters analysis about the weight mask on model training and data security. Also, the security analysis is limited to semi-honest settings, but it is hard to ensure in a real-world scenario. We will continue to improve this work from the perspective of algorithm design and then conduct a comprehensive analysis of the effectiveness of SFA. We will enhance this work in the future to achieve good efficiency while keeping the data security and model performance in neural networks in VFL.

\bibliographystyle{named}
\bibliography{ijcai22}

\begin{thebibliography}{}

\bibitem[\protect\citeauthoryear{Bonawitz \bgroup \em et al.\egroup
  }{2017}]{secure_agg}
Keith Bonawitz, Vladimir Ivanov, Ben Kreuter, Antonio Marcedone, H~Brendan
  McMahan, Sarvar Patel, Daniel Ramage, Aaron Segal, and Karn Seth.
\newblock Practical secure aggregation for privacy-preserving machine learning.
\newblock In {\em proceedings of the 2017 ACM SIGSAC Conference on Computer and
  Communications Security}, pages 1175--1191, 2017.

\bibitem[\protect\citeauthoryear{Chang and Lin}{2011}]{libsvm}
Chih-Chung Chang and Chih-Jen Lin.
\newblock Libsvm: a library for support vector machines.
\newblock {\em ACM transactions on intelligent systems and technology (TIST)},
  2(3):1--27, 2011.

\bibitem[\protect\citeauthoryear{Cheng \bgroup \em et al.\egroup }{2021a}]{sbt}
Kewei Cheng, Tao Fan, Yilun Jin, Yang Liu, Tianjian Chen, Dimitrios
  Papadopoulos, and Qiang Yang.
\newblock Secureboost: A lossless federated learning framework.
\newblock {\em IEEE Intelligent Systems}, 36(6):87--98, 2021.

\bibitem[\protect\citeauthoryear{Cheng \bgroup \em et al.\egroup
  }{2021b}]{haflo}
Xiaodian Cheng, Wanhang Lu, Xinyang Huang, Shuihai Hu, and Kai Chen.
\newblock Haflo: Gpu-based acceleration for federated logistic regression.
\newblock {\em arXiv preprint arXiv:2107.13797}, 2021.

\bibitem[\protect\citeauthoryear{Fu \bgroup \em et al.\egroup
  }{2022}]{label_attack}
Chong Fu, Xuhong Zhang, Shouling Ji, Jinyin Chen, Jingzheng Wu, Shanqing Guo,
  Jun Zhou, Alex~X Liu, and Ting Wang.
\newblock Label inference attacks against vertical federated learning.
\newblock In {\em 31st USENIX Security Symposium (USENIX Security 22)}, Boston,
  MA, August 2022. USENIX Association.

\bibitem[\protect\citeauthoryear{Hardy \bgroup \em et al.\egroup }{2017}]{vlr}
Stephen Hardy, Wilko Henecka, Hamish Ivey-Law, Richard Nock, Giorgio Patrini,
  Guillaume Smith, and Brian Thorne.
\newblock Private federated learning on vertically partitioned data via entity
  resolution and additively homomorphic encryption.
\newblock {\em arXiv preprint arXiv:1711.10677}, 2017.

\bibitem[\protect\citeauthoryear{Hu \bgroup \em et al.\egroup }{2019}]{fdml}
Yaochen Hu, Di~Niu, Jianming Yang, and Shengping Zhou.
\newblock Fdml: A collaborative machine learning framework for distributed
  features.
\newblock In {\em Proceedings of the 25th ACM SIGKDD International Conference
  on Knowledge Discovery \& Data Mining}, pages 2232--2240, 2019.

\bibitem[\protect\citeauthoryear{Kairouz \bgroup \em et al.\egroup
  }{2021}]{advanceFL}
Peter Kairouz, H~Brendan McMahan, Brendan Avent, Aur{\'e}lien Bellet, Mehdi
  Bennis, Arjun~Nitin Bhagoji, Kallista Bonawitz, Zachary Charles, Graham
  Cormode, Rachel Cummings, et~al.
\newblock Advances and open problems in federated learning.
\newblock {\em Foundations and Trends{\textregistered} in Machine Learning},
  14(1--2):1--210, 2021.

\bibitem[\protect\citeauthoryear{Lang}{1995}]{news20}
Ken Lang.
\newblock Newsweeder: Learning to filter netnews.
\newblock In {\em Proceedings of the Twelfth International Conference on
  Machine Learning}, pages 331--339, 1995.

\bibitem[\protect\citeauthoryear{LeCun \bgroup \em et al.\egroup
  }{2012}]{linear_init}
Yann~A LeCun, L{\'e}on Bottou, Genevieve~B Orr, and Klaus-Robert M{\"u}ller.
\newblock Efficient backprop.
\newblock In {\em Neural networks: Tricks of the trade}, pages 9--48. Springer,
  2012.

\bibitem[\protect\citeauthoryear{Luo \bgroup \em et al.\egroup
  }{2021}]{feature_inference_attack}
Xinjian Luo, Yuncheng Wu, Xiaokui Xiao, and Beng~Chin Ooi.
\newblock Feature inference attack on model predictions in vertical federated
  learning.
\newblock In {\em 2021 IEEE 37th International Conference on Data Engineering
  (ICDE)}, pages 181--192. IEEE, 2021.

\bibitem[\protect\citeauthoryear{Mahendran and
  Vedaldi}{2015}]{understanding_image}
Aravindh Mahendran and Andrea Vedaldi.
\newblock Understanding deep image representations by inverting them.
\newblock In {\em Proceedings of the IEEE conference on computer vision and
  pattern recognition}, pages 5188--5196, 2015.

\bibitem[\protect\citeauthoryear{Paillier}{1999}]{paillier}
Pascal Paillier.
\newblock Public-key cryptosystems based on composite degree residuosity
  classes.
\newblock In {\em Advances in Cryptology—EUROCRYPT’99}, 1999.

\bibitem[\protect\citeauthoryear{Vepakomma \bgroup \em et al.\egroup
  }{2018}]{split_learning_for_health}
Praneeth Vepakomma, Otkrist Gupta, Tristan Swedish, and Ramesh Raskar.
\newblock Split learning for health: Distributed deep learning without sharing
  raw patient data.
\newblock {\em arXiv preprint arXiv:1812.00564}, 2018.

\bibitem[\protect\citeauthoryear{Xiao \bgroup \em et al.\egroup
  }{2017}]{fmnist}
Han Xiao, Kashif Rasul, and Roland Vollgraf.
\newblock Fashion-mnist: a novel image dataset for benchmarking machine
  learning algorithms.
\newblock {\em arXiv preprint arXiv:1708.07747}, 2017.

\bibitem[\protect\citeauthoryear{Yang \bgroup \em et al.\egroup
  }{2019}]{flbook}
Qiang Yang, Yang Liu, Tianjian Chen, and Yongxin Tong.
\newblock Federated machine learning: Concept and applications.
\newblock {\em ACM Transactions on Intelligent Systems and Technology (TIST)},
  10(2):1--19, 2019.

\bibitem[\protect\citeauthoryear{Zhang \bgroup \em et al.\egroup
  }{2021}]{secure-bilevel}
Qingsong Zhang, Bin Gu, Cheng Deng, and Heng Huang.
\newblock Secure bilevel asynchronous vertical federated learning with backward
  updating.
\newblock In {\em Proceedings of the AAAI Conference on Artificial
  Intelligence}, volume~35, pages 10896--10904, 2021.

\bibitem[\protect\citeauthoryear{Zhou \bgroup \em et al.\egroup }{2018}]{din}
Guorui Zhou, Xiaoqiang Zhu, Chenru Song, Ying Fan, Han Zhu, Xiao Ma, Yanghui
  Yan, Junqi Jin, Han Li, and Kun Gai.
\newblock Deep interest network for click-through rate prediction.
\newblock In {\em Proceedings of the 24th ACM SIGKDD International Conference
  on Knowledge Discovery \& Data Mining}, pages 1059--1068, 2018.

\end{thebibliography}

\end{document}